\documentclass[prd, amsfonts, twocolumn, nofootinbib, showpacs]{revtex4}
\usepackage{graphicx, epsfig}
\usepackage{color}
\usepackage{amsmath}
\newcommand{\be}{\begin{equation}}
\newcommand{\ee}{\end{equation}}
\newcommand{\bea}{\begin{eqnarray}}
\newcommand{\eea}{\end{eqnarray}}

\newcommand{\gapp}{\mathrel{\raise.3ex\hbox{$>$}\mkern-14mu \lower0.6ex\hbox{$\sim$}}}
\newcommand{\lapp}{\mathrel{\raise.3ex\hbox{$<$}\mkern-14mu \lower0.6ex\hbox{$\sim$}}}
\def\bbox{{\,\lower0.9pt\vbox{\hrule \hbox{\vrule height 0.2 cm
\hskip 0.2 cm \vrule  height 0.2 cm}\hrule}\,}}

\begin{document}
\title{Analytic solution for a static black hole in the RSII model}
\author{De-Chang Dai and  Dejan Stojkovic}
\affiliation{Department of Physics,
SUNY at Buffalo, Buffalo, NY 14260-1500}


\begin{abstract}
We present here a static solution for a large black hole (whose horizon radius is larger than the AdS radius) located on the brane in RSII model. According to some arguments based on the AdS/CFT conjecture, a solution for the black hole located on the brane in RSII model must encode quantum gravitational effects and therefore can not be static. We demonstrated that a static solution can be found if the bulk is not empty. The stress energy tensor of the matter distribution in the bulk for the solution we found is physical (i.e. it is non-singular with the energy density and pressure not violating any energy conditions). The scale of the solution is given by a parameter "$a$".  For large values of the parameter "$a$" we have a limit of an almost empty AdS bulk.  It is interesting that the solution can not be transformed into the Schwarzschild-like form and does not reduce to the Schwarzschild solution on the brane. We also present two other related static solutions. At the end, we discuss why the numerical methods failed so far in finding static solutions in this context, including the solutions we found analytically here.
\widetext
\end{abstract}


\maketitle

\section{Introduction}

Despite an intensive search for a static solution of a large black hole located on the brane in the RSII model \cite{Randall:1999vf}, no such solution (analytic nor numeric) has been found so far. This lead some authors \cite{Tanaka:2002rb,Emparan:2002px}
to conjecture that such solutions do not exist at all. In the context of the AdS/CFT correspondence,
the black holes on a brane in an AdS$_5$ braneworld that solve the classical bulk equations with the brane as the boundary can be interpreted as duals of quantum-corrected
$4$-dimensional black holes, rather than classical ones, of a conformal field theory coupled to gravity.
According to this conjecture, solving the classical $(4+1)$-dimensional equations in the bulk is equivalent
to solving the $(3+1)$-dimensional equations, where the CFT stress-energy tensor includes the quantum effects of all planar diagrams.
These quantum effects incorporate the semiclassical Hawking radiation, vacuum polarization and perhaps other quantum effects.
The holographic dual theory with ${\cal O}(N)$ Yang-Mills fields may yield an effect proportional to $\sim N^2$, which would then persist in the large $N$ limit even as $\hbar \rightarrow 0$. Then such quantum radiation in the $4$-dimensional holographic theory would correspond to a classical $5$-dimensional process in the bulk, which would imply that a $5$-dimensional black hole localized on the brane should always classically radiate, and hence cannot be static.

This conjecture was disputed in \cite{Fitzpatrick:2006cd}, by arguing that this reasoning does not take into
account the strongly coupled nature of the holographic theory. It is unclear whether simply
multiplying the free field result by $N^2$ is valid when considering Hawking radiation. In particular, it may be that not all of the holographic degrees of freedom are available to the black hole of finite temperature since the number of asymptotic states that may be radiated is not enhanced by $N^2$ factor. In the case of the black hole localized on the brane, modes with mass smaller than the temperature have small overlap with the higher-curvature region of the brane black hole. Therefore only those modes
with mass of order the black hole temperature would be strongly coupled.
Similar argument may be made from another prospective. A black hole localized on the brane can not have any bulk components of angular momentum due to the bulk $Z_2$ symmetry (while there are no restrictions on the brane components of angular momentum). Then bulk radiation, which consists mostly of bulk Kaluza-Klein gravitons, must be highly suppressed \cite{Stojkovic:2004hp,Frolov:2002gf}, for the same reason as Schwarzschild black hole prefers to emit lower spin modes \cite{ddds}. Since presumably a large $N$ Yang-Mills fields on the CFT side correspond to a large number of bulk Kaluza-Klein gravitons on the AdS side, then radiation from such a black hole is not enhanced at all compared to the standard results.

The goal of this paper is not to unambiguously resolve the above conflict, but to present a static solution of a large black hole located on the brane in an AdS bulk with non-vanishing stress energy tensor. This may represent an important step toward the final resolution of this question.

We report here finding three different static solutions in this context. The first solution described by Eq.~(\ref{bh}) of Section \ref{1} is, as we believe, the most interesting. The horizon radius of the static black hole located on the brane in RSII model we found is larger than the AdS radius (which is a crucial requirement if we want the physics of the black hole to be affected by the non-trivial AdS bulk). The stress energy tensor of the matter distribution in the bulk for the solution we found is physical, i.e. not violating ant energy condition, and is everywhere non-singular. However, the solution can not be transformed into the Schwarzschild-like form and does not reduce to the Schwarzschild solution on the brane. In Section \ref{2}, we present a solution that reduces to the Schwarzschild-de Sitter-like black hole on the brane. However, the maximal value of the black hole horizon is of the order of the AdS radius for the physical bulk matter distribution. While such a black hole is not much smaller than the AdS radius, it does not probe the full AdS bulk either. Finally, if the AdS bulk is filled with a medium with the constant positive energy density there is another static solution, which we present in Section \ref{3}. The solution is not a black hole since it contains only a cosmological horizon, but it is interesting since the added bulk energy density does not change the brane tension which still depends solely on the negative AdS cosmological constant, which implies that the added bulk energy density does not necessarily cancel out completely the effects of the AdS space.
In the last Section \ref{ns}, we briefly discuss why the numerical methods failed so far in finding static solutions in this context, including the solutions we found analytically here.

\section{General conditions that a static solution has to satisfy}

We are looking for a static black hole which is spherically symmetric on the brane and axisymmetric in the bulk. We adopt the ansatz metric from \cite{Yoshino:2008rx,Kudoh:2003xz,Wiseman:2001xt}. The metric is

\begin{equation}
ds^2=\frac{\ell^2}{z^2}
\left[
-T^2dt^2+e^{2R}(dr^2+dz^2)+r^2e^{2C}d\Omega_2^2
\right],
\label{original-metric}
\end{equation}
where $d\Omega_2^2:=d\theta^2+\sin^2\theta d\phi^2$. Here, $\ell$ is related to the bulk cosmological constant as $\Lambda=-6/\ell^2$. The functions $T$, $R$, and $C$ depend only on $z$ and $r$. $r$ and $z$ can be transformed into $\rho$ and $\chi$ by

\begin{eqnarray} \label{coor}
r&=&\rho \sin \chi ,\\
z&=&\ell +\rho \cos\chi .
\end{eqnarray}
The location of the brane is at $\chi=\pi/2$.
The location of the event horizon is at $\rho=\rho_h$.

Define now a new tensor:
\begin{equation}
\mathcal{G}_{\mu\nu}:=R_{\mu\nu}-(2/3)\Lambda g_{\mu\nu}=0.
\end{equation}
$R_{\mu\nu}$ is Rieman tensor. The energy density without the bulk negative cosmology constant can be written as
\begin{equation}
G_{\mu\nu}=\mathcal{G}_{\mu\nu}-\frac{g_{\mu\nu}}{2}\mathcal{G}_{\alpha}^{\alpha}
\end{equation}

\begin{eqnarray}
\mathcal{G}^{t}_{t}&=&\frac{-z^2}{\ell^2 T e^{2R}}(\nabla^2T
+2\left(C_{,\rho}+\frac{2\ell}{z\rho}-\frac{1}{\rho}\right)T_{,\rho}\nonumber\\
&+&\frac{2}{\rho^2}\left(\cot\chi+C{,\chi}+\frac{2\rho}{z}\sin\chi\right)T_{,\chi}\nonumber\\
&+&\frac{2}{\rho z}\left(\sin\chi C_{,\chi}-\cos\chi\rho C_{,\rho}\right)T\nonumber\\
&+&\frac{4}{z^2}\left(1+\frac{\Lambda \ell^2}{6}e^{2R}\right)T )
\label{E1}
\end{eqnarray}

\begin{eqnarray}
\mathcal{G}^{t}_{t}&-&\mathcal{G}^{\rho}_{\rho}-\mathcal{G}^{\chi}_{\chi}+2\mathcal{G}^{ \theta}_{\theta}=\frac{2z^2}{\ell^2 e^{2R}}(\nabla^2 R -\frac{1-e^{2(R-C)}}{\rho^2\sin^2\chi}\nonumber\\
 &-&\frac{2}{z^2}\left(1+\frac{\Lambda \ell^2}{6}e^{2R}\right)
-\frac{2T_{,\rho}}{T}\left(C_{,\rho}+\frac{\ell}{z\rho}\right)\nonumber\\
 &-&\frac{2T_{,\chi}}{\rho^2 T}\left(\cot\chi+C_{,\chi}+\frac{\rho}{z}\sin\chi\right)-C_{,\rho}\left(C_{,\rho}+\frac{4\ell}{z\rho}-\frac{2}{\rho}\right)\nonumber\\
 &-&\frac{C_{,\chi}}{\rho^2}\left(C_{,\chi}+2\cot\chi+\frac{4\rho}{z}\sin\chi\right) )
 \label{E2}
\end{eqnarray}

\begin{eqnarray}
\mathcal{G}^{\theta}_{\theta}&=&\frac{-z^2}{\ell^2 e^{2R}}(\nabla^2 C+\frac{1-e^{2(R-C)}}{\rho^2\sin^2\chi} +\frac{4}{z^2}\left(1+\frac{\Lambda \ell^2}{6}e^{2R}\right)\nonumber\\
&+&\frac{T_{,\rho}}{T}\left(C_{,\rho}+\frac{\ell}{z\rho}\right)
+\frac{T_{,\chi}}{\rho^2 T}\left(\cot\chi+C_{,\chi}+\frac{\rho}{z}\sin\chi\right)\nonumber\\
&+&C_{,\rho}\left(2C_{,\rho}+\frac{5\ell}{z\rho}-\frac{1}{\rho}\right)\nonumber\\
&+&\frac{C_{,\chi}}{\rho^2}\left(2C_{,\chi}+4\cot\chi+\frac{5\rho}{z}\sin\chi\right) ).
\label{E3}
\end{eqnarray}

\begin{eqnarray}
\mathcal{G}^{\chi}_{\rho}&=&\frac{-2z^2}{\ell^2\rho^2 e^{2R}}(\frac{1}{2T}
\left[T_{,\rho\chi}-R_{,\chi}T_{,\rho}-T_{,\chi}\left(R_{,\rho}+\frac{1}{\rho}\right)\right]\nonumber\\
&+&C_{,\rho\chi}+C_{,\rho}(\cot\chi+C_{,\chi})
+\frac{R_{,\chi}}{2}\left(\frac{1}{\rho}-2C_{,\rho}-\frac{3\ell}{z\rho}\right)\nonumber\\
&-&R_{,\rho}\left(C_{,\chi}+\cot\chi+\frac{3\rho}{2z}\sin\chi\right) )
\label{E4}
\end{eqnarray}

\begin{eqnarray}
\mathcal{G}^{t}_{t}&-&\mathcal{G}^{\rho}_{\rho}+\mathcal{G}^{\chi}_{\chi}+2\mathcal{G}^{ \theta}_{\theta}=\frac{-2z^2}{\ell^2\rho^2 e^{2R}}(\frac{1}{T}
[T_{,\chi\chi}\nonumber\\
&+&T_{,\chi}\left(2\cot\chi+2C_{,\chi}-R_{,\chi}+\frac{3\rho}{z}\sin\chi\right)\nonumber\\
&+&\rho T_{,\rho}\left(2\rho C_{,\rho}+\rho R_{,\rho}+\frac{3\ell}{z}\right)
]+2C_{,\chi\chi}\nonumber\\
&+&C_{,\chi}\left(6\cot\chi+3C_{,\chi}+\frac{6\rho}{z}\sin\chi\right)\nonumber\\
&+&\rho C_{,\rho}\left(2\rho R_{,\rho}+\rho C_{,\rho}-2+\frac{6\ell}{z}\right)\nonumber\\
&-&R_{,\chi}\left(2\cot\chi+\frac{3\rho}{z}\sin\chi+2C_{,\chi}\right)+\rho R_{,\rho}\left(\frac{3\ell}{z}-1\right)\nonumber\\
&+&\frac{1-e^{2(R-C)}}{\sin^2\chi}
+\frac{6\rho^2}{z^2}\left(1+\frac{\Lambda\ell^2}{6}e^{2R}\right) )
\label{E5}
\end{eqnarray}

Here, $\nabla^2:=\partial_\rho^2+\partial_\rho/\rho+\partial_\chi^2/\rho^2$.
For a vacuum solution, the right hand sides of the above equations must be zero. However, we are looking for a solution with an energy distribution in the bulk, so the right hand sides, with an exception of the equation (\ref{E4}),  are not necessarily zero. Also, because of the energy distribution in the bulk, the spacetime does not necessarily reduce to the AdS spacetime at large distances. Therefore, we do not require $T=1$, $C=0$, and $R=0$ at large distances. We believe that this is a very important assumption. Namely, the empty AdS bulk does not seem to allow for a static solution. In order to cancel destabilizing effects of the AdS bulk, it appears that one needs to fill the AdS bulk with some sort of matter/energy distribution. The effects of the negative AdS cosmological constant are small or negligible along the brane (at distances smaller than the AdS radius), while they are strong at distances larger than the AdS radius. Extra bulk energy distribution that dies off with the distance from the brane would have satisfied the AdS boundary conditions ($T=1$, $C=0$, and $R=0$ at large distances), however it is very unlikely that it would stabilize the black hole solution. Therefore we do not require $T=1$, $C=0$, and $R=0$ at large distances.

We consider the brane to be the vacuum outside the horizon. Israel's junction condition then implies that
\begin{equation}
\frac{T_{,\chi}}{T}=R_{,\chi} = C_{,\chi} =\frac{\rho}{\ell}(e^R-1)
\label{con-1}
\end{equation}

For self-consistency, we assume that there is no singular mass distribution on the symmetry axis, $\chi=0$. The regularity conditions are:

\begin{eqnarray} \label{con2}
T_{,\chi}=R_{,\chi} = C_{,\chi} =0\\
R=C \nonumber
\end{eqnarray}

On the black hole horizon, we have $\rho =\rho_h$, $T=0$. The regularity of equations (\ref{E2}) to (\ref{E5}) implies

\begin{eqnarray}
\label{con-3}
C_{,\rho}&=&-\frac{\ell}{z\rho}\\
T_{,\rho\chi}&=&R_{,\chi}T_{,\rho}\\
R_{,\rho}&=&-\frac{\ell}{z\rho}
\end{eqnarray}
These boundary conditions are sufficient but not necessary. If they are satisfied, the energy density will not diverge near horizon.
Since the equations we need to solve are very complicated, we will look for a special solution for which

\begin{eqnarray} \label{rec}
&& \frac{T_{,\chi}}{T}=R_{\chi}\\
&& R=C \nonumber
\end{eqnarray}
is satisfied over the whole spacetime (not only at the horizon).  This new requirement does not violate the boundary conditions.

We now substitute (\ref{rec}) into (\ref{E4}), $\mathcal{G}^{\chi}_{\rho}=0$, which is now simplified to
\begin{equation}
R_{,\chi\rho}-R_{,\chi}R_{,\rho}-\frac{\ell R_{,\chi}}{z\rho}-\frac{R_\rho \rho \sin\chi}{z}=0
\end{equation}

This equation has the general solution

\begin{equation}
R_{,\chi}=\frac{\rho}{z}(e^R f(\chi)-\sin\chi)
\end{equation}

Here $f(\chi)$ is a function which depends on $\chi$ only.  $T$, $R$, and $C$ can then be found
\begin{eqnarray}
e^R=e^C&=&\frac{z}{g(\rho)-\rho F(\chi)}\\
T&=&\frac{h(\rho) z}{g(\rho)-\rho F(\chi)}
\end{eqnarray}
Here, $F,_{\chi}=f$. $h(\rho)$ and $g(\rho)$ are functions that depend on $\rho$ only. To satisfy (\ref{con-1}) and (\ref{con2}), we must have $f(\frac{\pi}{2})=1$ and $f(0)=0$ respectively. We thus  choose $f(\chi)=\sin\chi$, and the above equations become

\begin{eqnarray}
e^R=e^C&=&\frac{z}{g(\rho)+\rho \cos\chi}\\
T&=&\frac{h(\rho) z}{g(\rho)+\rho \cos\chi}
\end{eqnarray}

To satisfy equation (\ref{con-3}), we must have
\begin{equation}
g=\rho g_{,\rho}
\label{bound-1}
\end{equation}
at $\rho=\rho_h$. Also, to regularize equation (\ref{E1}) at $\rho =\rho_h$, we write
\begin{equation}
T_{,\rho\rho}+(2\frac{\ell}{z\rho}-\frac{1}{\rho})T_{,\rho}=0.
\end{equation}

$h(\rho)$ has to satisfy

\begin{eqnarray}
\label{bound-2}
h&=&0\\
h_{,\rho}&=&\rho h_{,\rho\rho}
\label{bound-3}
\end{eqnarray}
at $\rho=\rho_h$. Again, this regularity condition is sufficient but not necessary.

Following the established procedure, we can find the surface temperature of a black hole with the horizon radius $\rho_h$ as
\begin{equation}\label{sg}
\kappa = -\frac{1}{2}\sqrt{\frac{g^{\rho\rho}}{-g_{tt}}}\frac{\partial g_{tt}}{\partial \rho}|_{\rho=\rho_h}=e^{-R}T_{,\rho} |_{\rho=\rho_h}=h_{,\rho}(\rho_h)
\end{equation}

The functions $g(\rho)$ and $h(\rho)$ uniquely define the metric.
We now have a set of conditions (\ref{bound-1}), (\ref{bound-2}) and (\ref{bound-3}) that these functions have to satisfy, and we are going to explore if there is a reasonable solution in the next section.

\section{Particular Solutions}

\subsection{Static black hole solution larger than the AdS radius}
\label{1}

It is not very difficult to find a solution which satisfies all of the conditions. What is difficult is to find an appropriate solution whose corresponding stress energy tensor is physical (i.e. non-singular and not violating energy conditions). For example, in \cite{Kanti:2001cj}, a good progress has been made, but the procedure lead to non-physical solutions.

We first try the following solution for $g(\rho)$ and $h(\rho)$ that satisfies all the conditions:
\begin{eqnarray}
h(\rho)&=&\rho^2/\rho_h^2-1,\\
\label{ch-1}
g(\rho)&=&\ell\rho/a.
\end{eqnarray}
This choice uniquely defines the metric.
Here $\rho_h$ is the horizon radius, while $a$ is a constant with units of length which determines the scale of the solution. The explicit metric is
\begin{eqnarray} \label{bh}
ds^2&=&-\frac{\ell^2}{\rho^2}\left(\frac{\frac{\rho^2}{\rho_h^2}-1 }{(\frac{\ell}{a})+ \cos\chi}\right)^2dt^2+\frac{\ell^2}{\rho^2}\left(\frac{1}{(\frac{\ell}{a})+ \cos\chi}\right)^2\nonumber\\
 &\times &(d\rho^2+\rho^2 d \chi^2 +\rho^2 \sin^2 \chi d\Omega^2_2)
\end{eqnarray}

The solution is given in isotropic coordinates.
The correspondence between the coordinate $\rho$ and a usual radial coordinate $r$ is given by (\ref{coor}). The brane is located at $\chi =\pi/2$, thus $r=\rho$ on the brane.  The true physical singularity is located at $\rho=0$, while the event horizon is at $\rho=\rho_h$.
The horizon is spherically symmetric in $\rho$ coordinate, but it is squashed in the direction of extra dimension in the $r$ coordinate. Thus, the solution looks like a pancake rather than a cigar, since it extension along the brane is larger than the extension in the bulk direction. This is reasonable since the cigar-like solution will be unstable, suffering from the Gregory-Laflamme instability.

From Eq.~(\ref{sg}), we find the surface gravity of this black hole
\begin{equation}
\label{surface-1}
\kappa=\frac{2}{\rho_h}
\end{equation}

However, we see that the solution does not reduce to the Schwarzschild solution on the brane! It is impossible to perform coordinate transformation to put this solution in a Schwarzschild-like form.

The stress-energy distribution corresponding to this solution (after subtracting the negative cosmological constant of the original AdS space) is
\begin{eqnarray}\label{type-1}
G^t_t&=&\frac{3(\frac{a}{\ell}\cos\chi -1)}{a^2}\\
G^\chi_\chi&=&\frac{6\cos\chi}{a\ell} \nonumber \\
G^r_r &=& G_t^t  \nonumber \\
G^\theta_\theta &=& G^\phi_\phi = G^\chi_\chi  \nonumber
\end{eqnarray}

For a physically acceptable solution we need to satisfy $G^t_t <0$ and $|G^t_t |\geq |G^\mu_\mu |$, so we must have $\ell>3a$. Such distribution does not violate any energy condition.
It is instructive to note that this bulk stress energy tensor is also regular at the AdS horizon, i.e. at $z=\infty$, as can be directly seen from Eq.~(\ref{type-1}). Since $\rho_h$ does not appear in the stress-energy distribution, $\rho_h > \ell$ can be freely set. Thus, this solution represents a static black hole located on the brane in RSII scenario, whose horizon radius is larger than the AdS radius.

The magnitude of the components of the stress energy tensor is given by the parameter $a$. For very large $a$, the bulk is almost empty AdS (we can not take an extreme $a \rightarrow \infty$ limit since in this case the metric (\ref{bh}) would have a branch-cut at $\chi =\pi/2$). Thus, the solution with $\rho_h > \ell> 3a$ for large values of $a$ is as close as we can get to the large static black hole solution in the empty AdS bulk.

In the light of the conjecture that static vacuum black hole solutions do not exist, the stress energy tensor in (\ref{type-1}) maybe interpreted as the energy needed to compensate for the black hole energy lost to quantum radiation. Note however that the stress energy tensor depends only on $a$ and $\ell$, and not on $\rho_h$, but the surface gravity (and therefore the Hawking temperature) of the black hole is given by an inverse $\rho_h$. Therefore, unless there is some non-trivial relation between $a$, $\ell$, and $\rho_h$, this is an unlikely explanation.

The other possibility is to interpret the bulk stress energy tensor as the property of the black hole itself. For example, a charged Reissner–Nordstrom black hole is itself a source of a non-zero stress energy tensor. The magnitude of the black hole charge would be then given by the parameter $a$ (and of course $\ell$), since both the metric (\ref{bh}) and the stress energy tensor (\ref{type-1}) depend on this parameter. In that case, the solution (\ref{bh}) represents a charged static black hole located on the brane in RSII scenario. For a possible Lagrangian describing the fields that give the appropriate stress energy tensor see Appendix \ref{app}.

At the end of this Section, we write down a metric similar to (\ref{bh}), which also supports the stress energy tensor (\ref{type-1}):
\begin{eqnarray}
ds^2&=&-\frac{\ell^2}{\rho^2}\left(\frac{\frac{\rho^2}{\rho_h^2} }{(\frac{\ell}{a})+ \cos\chi}\right)^2dt^2+\frac{\ell^2}{\rho^2}\left(\frac{1}{(\frac{\ell}{a})+ \cos\chi}\right)^2\nonumber\\
 &\times &(d\rho^2+\rho^2 d \chi^2 +\rho^2 \sin^2 \chi d\Omega^2_2)
\end{eqnarray}
For this object, the physical singularity and the horizon (which is in isotropic coordinates given by $g_{00}=0$) coincide at $\rho =0$.
This is an interesting situation since the singularity at $\rho =0$ can not be reached in finite time as seen by an outside observer.

\subsection{Solution that reduces to the Schwarzschild-de Sitter-like black hole on the brane}
\label{2}

In this Section we will look for the solution that reduces to the Schwarzschild-de Sitter-like black hole on the brane.
We try the following solution for $g(\rho)$ and $h(\rho)$:

\begin{eqnarray}
\label{c3}
\frac{r}{\rho}&=&\frac{\frac{dr}{d\rho}}{\sqrt{1-\frac{M^2}{r^2}-\frac{r^2}{H^2}-\frac{r}{A}}}=\frac{\ell}{g(\rho)}\\
\label{c3-1}
\frac{h(\rho)\ell}{g(\rho)}&=&\sqrt{1-\frac{M^2}{r^2}-\frac{r^2}{H^2}-\frac{r}{A}}
\end{eqnarray}
where $M$, $H$ and $A$ are constants.
By integrating equation (\ref{c3}), one finds
\begin{equation}\label{rrho}
r=E(\rho)
\end{equation}
where $E(\rho)$ stands for a kind of an elliptic integral of the variable $\rho$.
From equation (\ref{c3}) and (\ref{c3-1}), the functions $g(\rho)$ and $h(\rho)$ can be written as

\begin{eqnarray}
g(\rho )&=&\frac{\ell\sqrt{1-\frac{M^2}{E(\rho )^2}-\frac{E(\rho)^2}{H^2}-\frac{E(\rho)}{A}}}{\frac{dE(\rho )}{d\rho}}\\
h(\rho )&=&\frac{1-\frac{M^2}{E(\rho )^2}-\frac{E(\rho)^2}{H^2}-\frac{E(\rho)}{A}}{\frac{dE(\rho )}{d\rho}}\\
\end{eqnarray}

The explicit metric is

\begin{eqnarray}
ds^2&=&-\left(\frac{\frac{1-\frac{M^2}{E(\rho )^2}-\frac{E(\rho)^2}{H^2}-\frac{E(\rho)}{A}}{\frac{dE(\rho )}{d\rho}} }{\frac{\sqrt{1-\frac{M^2}{E(\rho )^2}-\frac{E(\rho)^2}{H^2}-\frac{E(\rho)}{A}}}{\frac{dE(\rho )}{d\rho}}+ \frac{\rho\cos\chi}{\ell}}\right)^2dt^2\nonumber \\
&+&\left(\frac{1}{\frac{\sqrt{1-\frac{M^2}{E(\rho )^2}-\frac{E(\rho)^2}{H^2}-\frac{E(\rho)}{A}}}{\frac{dE(\rho )}{d\rho}}+ \frac{\rho\cos\chi}{\ell}}\right)^2\nonumber\\
 &\times &(d\rho^2+\rho^2 d \chi^2 +\rho^2 \sin^2 \chi d\Omega^2_2)
\end{eqnarray}

The metric on the brane reduces to

\begin{eqnarray}
ds^2&=&-\left(1-\frac{M^2}{r^2}-\frac{r^2}{H^2}-\frac{r}{A}\right)dt^2\nonumber\\
&+&\left(1-\frac{M^2}{r^2}-\frac{r^2}{H^2}-\frac{r}{A}\right)^{-1}dr^2+r^2d\Omega_2^2
\end{eqnarray}

For large enough $A$, this is the metric of a black hole inside de-Sitter space. $H$ is the positive cosmological constant. The black hole horizon is within the de-Sitter horizon. Since the solution is not exactly Schwarzschild-de Sitter because of the term $r/A$, the space between the two horizons contains some other contribution besides the positive cosmological constant. The energy density can be calculated from the Einstein tensor.

\begin{eqnarray}
G^t_t&=&\frac{-12rA-9H^2}{2rH^2A}+\frac{12M^2A+3r^3}{2r^3A\ell}\cos\chi\\
G^\chi_\chi&=&\frac{-6rA-3H^2}{rH^2A}+\frac{3\cos\chi}{A\ell}-\frac{6M^2\cos^2\chi}{r^2\ell^2}\\
G^r_r &=& G_t^t\\
G^\theta_\theta &=& G^\phi_\phi = G^\chi_\chi
\end{eqnarray}

If we remove the positive cosmological constant contribution the remaining matter is

\begin{eqnarray}
T^t_t=G^t_t-G^\chi_\chi &=&\frac{-3}{2rA}+\frac{12M^2A-3r^3}{2r^3A\ell}\cos\chi\nonumber\\
   &+&\frac{6M^2\cos^2\chi}{r^2\ell^2}\\
T^r_r &=& T_t^t\\
T^\theta_\theta &=& T^\phi_\phi = T^\chi_\chi=0
\end{eqnarray}

A reasonable physical solution requires $T^t_t<0$. This condition is easy to satisfy by choosing $A>0$ and a large $\ell$.
However, large $\ell$ implies that it is very difficult to have a black hole horizon larger than AdS radius. In fact, one can show that the maximal value of the black hole horizon is $r_h =\ell/2$. While such a black hole is not much smaller than the AdS radius, it does not probe the full AdS bulk either.

From Eq.~(\ref{sg}), we find the surface gravity of this black hole
\begin{equation}
\kappa=\frac{M^2}{r_h^3}-\frac{r_h}{H^2}-\frac{1}{2A}
\end{equation}
where the relation between $r_h$ and $\rho_h$ is given by Eq.~(\ref{rrho}).

\subsection{Static solution with the cosmological horizon}
\label{3}

In this Section we present one more static solution.
We try the following solution that also satisfies all the conditions:
\begin{eqnarray}
h(\rho)&=&\rho^2/\rho_h^2-1,\\
g(\rho)&=&\ell (1+\rho^2/\rho_h^2).
\end{eqnarray}
It is easy to check that equations (\ref{bound-1}), (\ref{bound-2}), and (\ref{bound-3}) are satisfied. The metric is

\begin{eqnarray}
ds^2=&&\!\!\!\!\!\!-\!\!\left(\frac{\frac{\rho^2}{\rho_h^2}-1 }{(\frac{\rho^2}{\rho_h^2}+1)+\frac{\rho}{\ell} \cos\chi}\right)^2\!\!dt^2+\left(\frac{1}{(\frac{\rho^2}{\rho_h^2}+1)+\frac{\rho}{\ell} \cos\chi}\right)^2\nonumber\\
 &&\times \left(d\rho^2+\rho^2 d \chi^2 +\rho^2 \sin^2 \chi d\Omega^2_2\right)
\end{eqnarray}

The energy density that corresponds to this solution is
\begin{equation}
G^{t}_{t}=G^{\rho}_{\rho}=G^{\chi}_{\chi}=G^{\theta}_{\theta}=G^{\phi}_{\phi}=-24/\rho_h^2
\end{equation}

This energy density corresponds to a positive cosmological constant in bulk. The horizon at $\rho_h$ is not a black hole horizon; it is de-Sitter horizon. The region that is of interest is $\rho_h>\rho>0$, since in these coordinates, the region $\rho>\rho_h$ is identical to $\rho_h>\rho>0$.

The AdS space itself has the negative bulk cosmology constant, $\Lambda$, which can be now partially canceled by the positive cosmological constant of the bulk ``dark energy" distribution. However, the solution is still interesting because the brane tension still depends on the AdS $\Lambda$ only. The bulk dark energy density does not change the brane tension.

A special limit will be $\rho_h =2 \ell$, when the positive cosmological constant completely cancels out $\Lambda$, and the net bulk cosmological constant is zero. The horizon still exists even in this case, but this is just an original horizon of the tense brane which was removed by the fine tuned negative AdS cosmological constant in the RSII scenario. Addition of the positive cosmological constant canceled out the contribution of the AdS cosmological constant and the horizon reappears.

\section{Possible problem with Numerical Methods for finding the solutions}
\label{ns}

\begin{figure}[t]
   \centering
\includegraphics[width=3.2in]{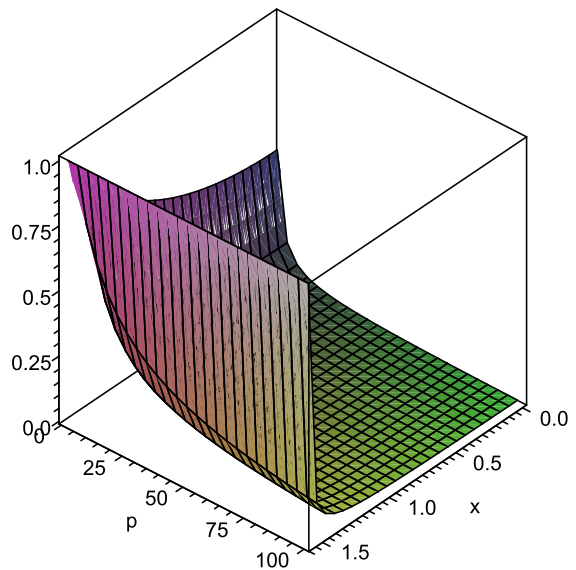}
\caption{ $\ell/z$ as a function of $\rho$ and $\chi$. $\ell/z$ changes very fast in $\chi$ direction at large $\rho$, behaving effectively as the step function. This makes the numerical methods very likely to fail.}
    \label{z-1}
\end{figure}

It will be useful to try to look for solutions of the problem of a large static black hole in RSII model using numerical methods. However, numerical methods can not efficiently deal with the step functions or discontinuous regions. Equations (\ref{E1}) to (\ref{E5}) include the term $1/\sin^2\chi$, which causes numerical instabilities near the axis of symmetry. This fact is well known, and indeed  $1/\sin^2\chi$ term can be dealt with by some numerical techniques. However, there is another term, $\ell/z$, which causes similar problems. $\ell/z$ becomes very close to the step function in $\chi$ direction at large distance. Figure \ref{z-1} shows how $\ell/z$ changes with $\rho$ and $\chi$. Obviously, $\ell/z$ changes very fast in $\chi$ direction at large $\rho$. Therefore, solving the problem numerically will require the resolution in $\chi$ direction to increase with distance. However, the resolution in $\chi$ direction does not become better as the distance becomes larger \cite{Yoshino:2008rx,Kudoh:2003xz}.
We believe that this is one of the reasons why numerical methods failed so far in finding static solutions, including the solutions we found analytically here.

\section{Conclusions}

We presented here a static solution for a large black hole (whose horizon radius is larger than the AdS radius) located on the brane in RSII model. According to some arguments coming from the AdS/CFT conjecture, a solution for the black hole located on the brane in RSII model must encode quantum gravitational effects and therefore can not be static. We demonstrated that a static solution can be found if there is a non-zero bulk matter distribution. The stress energy tensor of the matter distribution in the bulk for the solution we found is physical (i.e. non-singular and it does not violate any energy condition). The scale of the solution is given by a parameter "$a$".  For large values of the parameter "$a$" we have a limit of an almost empty AdS bulk (in which case the CFT dual is clear). While this does not directly disprove the conjecture made in Ref.~\cite{Tanaka:2002rb,Emparan:2002px}, it comes as close as currently possible. It is interesting that the solution can not be transformed into the Schwarzschild-like form and does not reduce to the Schwarzschild solution on the brane.

We also presented a solution that reduces to the Schwarzschild-de Sitter-like black hole on the brane. However, the maximal value of the black hole horizon is of the order of the AdS radius for the physical bulk matter distribution. While such a black hole is not much smaller than the AdS radius, it does not probe the full AdS bulk either. Finally, if the AdS bulk is filled with dark energy, i.e. a medium with the constant positive energy density there is another static solution. The solution is not a black hole since it contains only a cosmological horizon, but it is interesting since the added bulk energy density does not change the brane tension which still depends solely on the negative AdS cosmological constant (i.e. the added bulk energy density does not necessarily cancel out completely the effects of the AdS space).
We also discuss why the numerical methods failed so far in finding static solutions, including the solutions we found analytically here.

The non-zero bulk stress energy tensor distribution that stabilizes the black hole in the RSII setup can be interpreted in two ways. The first possibility is that the bulk must be heated in order to compensate for the energy lost by a black hole. We find this interpretation unlikely since the surface gravity (and thus the temperature) of the black hole depends on the horizon radius $\rho_h$, while the bulk matter distribution that stabilizes the black hole does not depend on $\rho_h$. The second possibility is to interpret the bulk stress energy tensor as the property of a black hole itself (in analogy with the charged Reissner–Nordstrom black hole which is itself a source of a non-zero stress energy tensor).
In that case, the solution (\ref{bh}) represents a charged static black hole located on the brane in RSII scenario (for related work see \cite{arXiv:1103.4758,arXiv:1103.3352,arXiv:1101.1384,arXiv:1003.2572}).

Another area of interest for the black hole solutions presented here are searches for potential violations of the no-hair theorems in astrophysical black holes \cite{arXiv:1105.5645,arXiv:1105.3191}, since the metric in Eq.~({\ref{bh}) does not reduce to the  Schwarzschild solution on the brane.

\section{Appendix}
\label{app}

We give here a possible Lagrangian which describes the fields that give the appropriate stress energy tensor in Eq.~(\ref{type-1}).
Consider the following Lagrangian
\begin{equation} \label{L}
L=\int dx^5 \left[ -f(\chi) F^{\mu \nu}F_{\mu \nu}+V(\chi) \right]
\end{equation}
where $F_{\mu \nu}$ is the field strength for some vector field $A_\mu$, and $f(\chi)$ is an arbitrary function of the coordinate $\chi$. The Lagrangian is not Lorentz invariant from a five-dimensional point of view since it depends explicitly on the coordinate $\chi$. However, for an observer located on the brane (where $\chi = \pi/2$) the Lagrangian is just that of a vector field. Since the presence of the brane in the RSII model breaks the full five-dimensional Lorentz invariance anyway, requiring the full five-dimensional Lorentz invariance for the Lagrangian giving an appropriate stress energy tensor seems unreasonable.

The energy momentum tensor for the Lagrangian in Eq.~(\ref{L}) is

\begin{equation}
T^\mu_\nu =f(\chi)(4F^{\mu \alpha} F_{\nu \alpha}-g^\mu_\nu F^{\alpha \beta}F_{\alpha \beta})- V(\chi)
\end{equation}

The equation of motion of $A^{\mu}$ is then

\begin{equation}
\label{constraint}
(f(\chi )F^{\mu \nu})_{;\mu}=0
\end{equation}

Consider the $A_{\mu}$ in the form

\begin{equation}
\label{vector}
A_\mu=(A_0(\rho),0,0,0)
\end{equation}

If we insert Eq.~(\ref{vector}) into Eq.~(\ref{constraint}) we get
\begin{equation}
\frac{1}{\sqrt{-g}}\partial_\rho \left(\sqrt{-g}f(\chi)F^{\rho t}\right)=0
\end{equation}
One then has
\begin{equation}
F^{\rho t} = \frac{b \rho^2 (\frac{\l}{a}+\cos\chi)^4}{\l^2\left(\frac{\rho^2}{\rho^2_h}-1\right)}
\end{equation}
where $b$ is a constant. The the stress energy tensor is then
\begin{eqnarray}
T^t_t&=&-2b^2(\frac{\l}{a}+\cos\chi)^4f(\chi)-V(\chi)\\
T^\chi_\chi &=& 2b^2(\frac{\l}{a}+\cos\chi)^4f(\chi)-V(\chi)\\
T^\rho_\rho & =&T^t_t\\
T^\theta_\theta &=&T^\phi_\phi =T^\chi_\chi
\end{eqnarray}

For a specific choice of $2b^2(\frac{\l}{a}+\cos\chi)^4f(\chi)=\frac{3(\frac{a}{\l}\cos\chi+1)}{2a\l}$ and $V(\chi)=\frac{-3(3\frac{a}{\l}\cos\chi-1)}{2a\l}$, the stress energy tensor is exactly the same as in Eq.~(\ref{type-1}).

\begin{acknowledgments}
The authors would like to thank N. Arkani-Hamed for suggesting to look for these solution in a non-empty AdS bulk and N. Kaloper for useful discussions. D.S. acknowledges the financial support from NSF, grant number PHY-0914893.

\end{acknowledgments}

\end{document}